\documentstyle[twoside,fleqn]{article}
\newcommand{\bls}[1]{\renewcommand{\baselinestretch}{#1}}

\def\twocol{\twocolumn \mathindent=1em}
\def\noi{\noindent}

\newcommand{\psection}[1]{{\raggedright\section{\protect\large\bf #1}}}

\newcommand{\psubsection}[1]%
   {{\raggedright\subsection{\protect\normalsize\bf #1}}}

\renewcommand{\thesubsubsection}%
   {\arabic{section}.\arabic{subsection}.\arabic{subsubsection}.}
\newcommand{\heads}[2]{\markboth{\protect\small\it #1}{\protect\small\it #2}}
\newcommand{\Acknow}[1]{\subsection*{Acknowledgement} #1}

\newcommand{\Title}[1]{\noindent {\Large #1} \\}

\newcommand{\Authors}[4]{\noindent
     {\large\bf #1\dag\ #2\ddag}\medskip\begin{description}
     \item[\dag]{\it #3} \item[\ddag]{\it #4}\end{description}}

\newcommand{\foom}[1]{\protect\footnotemark[#1]}
\newcommand{\foox}[2]{\footnotetext[#1]{#2}}
\newcommand{\email}[2]{\footnotetext[#1]{e-mail: #2}}

\newcommand{\WFighere}[3]%           % Applicable only in one-column mode
       {\framebox[\textwidth]{\rule{0cm}{#2}} \vspace{-3pt} \\
        Figure #1: {\small #3 } \smallskip\hrule\bigskip
        \addtocounter{figure}{1}}

\newcommand{\Ref}[1]{Ref.\,\cite{#1}}

\newcommand{\sect}[1]{Sec.\,#1}

\def\nq{\hspace{-1em}}
\def\nqq{\hspace{-2em}}
\def\nhq{\hspace{-0.5em}}

\def\cm{\hspace{1cm}}

\def\eq{Eq.\,}
\def\eqs{Eqs.\,}
\def\beq{\begin{equation}}
\def\eeq{\end{equation}}
\def\bear{\begin{eqnarray}}
\def\al{&\nhq}
\def\lal{&&\nqq {}}               % left alignment
\def\bearr{\begin{eqnarray} \lal}
\def\ear{\end{eqnarray}}
\def\earn{\nonumber \end{eqnarray}}

\def\tst{\textstyle}

\def\nn{\nonumber\\ {}}

\def\nnn{\nonumber\\ \lal }

\def\yy{\\[5pt]}

\def\eql{\al =\al}

\def\e{{\,\rm e}}

\def\dim{{\,\rm dim\,}}

\def\half{{\tst\frac{1}{2}}}

\newcommand{\aver}[1]{\langle \, #1 \, \rangle \mathstrut}

\catcode`\@=11 \@addtoreset{equation}{section}\catcode`\@=12
\newcommand{\be}{\begin{equation}}
\newcommand{\ee}{\end{equation}}
\newcommand{\ba}{\begin{eqnarray}}
\newcommand{\ea}{\end{eqnarray}}
\newcommand{\p}{\partial}
\newcommand{\btd}{\bigtriangledown}
\newcommand{\btu}{\bigtriangleup}
\newcommand{\sq}[1]{\sqrt{|#1|}}
\newcommand{\avers}[1]{{\aver{#1}}_*}

\heads{V.D. Ivashchuk and V.N. Melnikov}
{Multidimensional Gravity with Einstein Internal Spaces}

\bls{1.01}
\begin{document}
\setcounter{page}{0}
\thispagestyle{empty}
\large
\begin{center}
               RUSSIAN GRAVITATIONAL SOCIETY\\
               INSTITUTE OF METROLOGICAL SERVICE \\
               CENTER OF GRAVITATION AND FUNDAMENTAL METROLOGY\\
\end{center}
\vskip 4ex
\begin{flushright}
                                         RGS-VNIIMS-003-96\\
                                         hep-th/9612054
\end{flushright}
\vskip 25mm
\begin{center}

\Title{\Large\bf  MULTIDIMENSIONAL GRAVITY \yy WITH
                  EINSTEIN INTERNAL  SPACES }

\bigskip

\Authors{V.D. Ivashchuk\foom 1} {and V.N. Melnikov\foom 2}
{Centre for Gravitation and Fundamental Metrology, VNIIMS,
     3--1 M. Ulyanovoy Str., Moscow 117313, Russia}
{CNRS, Centre de Physique Theorique, Luminy, Case 9-07,
-F13288, Marseille, CEDEX 9, France}

\end{center}

\bigskip
\noi     {\bf Abstract}
\medskip

\noi
A multidimensional gravitational model on the manifold
$M = M_0 \times \prod_{i=1}^{n} M_i$, where $M_i$  are
Einstein spaces ($i \geq 1$), is studied.
For $N_0  = \dim  M_0 > 2$
the $\sigma$ model representation is considered
and it is shown that the  corresponding Euclidean
Toda-like system  does not satisfy the Adler-van-Moerbeke
criterion. For  $M_0 = {\bf R}^{N_0}$, $N_0 = 3, 4, 6$ (and
the total dimension $D = \dim  M = 11, 10, 11$, respectively)
nonsingular spherically symmetric solutions to vacuum
Einstein equations are obtained and their generalizations to
arbitrary signatures are considered.
It is proved that for a non-Euclidean signature
the Riemann tensor squared of the solutions diverges
on certain hypersurfaces in ${\bf R}^{N_0}$.

\vfill

{\small \noi Submitted to {\it Gravitation and Cosmology}}
\vspace {1cm}

\centerline{Moscow 1996}]

\email 1 {ivas@cvsi.rc.ac.ru}
\foox 2 {On leave from
Centre for Gravitation and Fundamental Metrology, VNIIMS,
     3--1 M. Ulyanovoy Str., Moscow 117313, Russia}

\twocol
\normalsize

\psection{Introduction}

Our paper is devoted to studying a  model of
multidimensional gravity considered previously in Refs.\,[1--3]
(see also [24--27]).
This model contains ``our space" $ M_0$
of dimension $N_{0}$ and a set of internal Einstein
spaces $M_{1}, \ldots , M_{n}$. All scale factors of $M_i$
are supposed to be functions on $ M_0$. For physical applications
$N_{0} \leq 4$ (e.g $N_0 =1, 2$ corresponds to cosmology and
axial symmetry, respectively).

On the classical level the model is equivalent to some
tensor-multiscalar theory and may be also treated as a generalization
of the standard Brans-Dicke theory with the parameter $\omega =1/N' -1$,
where $N'$ is the total internal space dimension \cite{RZ}.

It should be noted that scalar-tensor theories are rather
popular now (see, for example \cite{SM}-\cite{A}).

For $N_0 {=} 1$  we get a multidimensional cosmological model
considered by  many authors \cite{CD}-\cite{GIKM}.
This model was reduced
to a pseudo-Euclidean Toda-like system \cite{IMZ,I,IM2,GIM1}, which  is
a rather nontrivial object, since there are no explicit integration
methods or integrability conditions when the number of spaces with nonzero
curvature is greater than one. Recently, in \Ref{GIM2} three integrable
non-trivial families of solutions were obtained for a cosmological
model with two nonzero curvatures ($n=2$) and $(N_{1} = \dim  M_{1},
N_{2} = \dim  M_{2}) = (3,6), (5,5), (2,8)$ by solving the Abel
equation. They include nonsingular
spherically symmetric solutions on manifolds ${\bf R}^{7} \times
M_2$  \cite{GIM2} and ${\bf R}^{6} \times M_2$ \cite{GIKM} for $\dim
M = 3 $ and $5$, respectively.  As it is  hard to solve the
Abel equation from \cite{GIM2} for arbitrary $(N_1, N_2)$, we may first try
to obtain nonsingular spherically symmetric solutions on
${\bf R}^{N_0} \times M_2$  and then try to extend them
to the general solution for a cosmology with two curvatures on the
manifold ${\bf R}_{+} \times {\bf S}^{N_0 -1} \times M_2$.

The paper is organized as follows. In \sect 2 we describe
the model and obtain the equations of motion. In \sect 3
the non-exceptional case  $N_0 \neq 2$  is considered. We
obtain a generalized $\sigma$ model and in the case $N_0 > 2$
(such that the ``midisuperspace" metric is Euclidean) show that the
interaction potential does not satisfy the Adler-van-Moerbeke
criterion \cite{AM}. We diagonalize the ``midisuperspace" metric
and obtaine a ``diagonalized" $\sigma$ model representation
in a more explicit manner than in \cite{Ber,RZ}.
In \sect 4 three families of nonsingular
spherically symmetric solutions with the topology
${\bf R}^{N_0} \times M_{1} \times \ldots \times M_{n}$
are obtained  for $N_0 = 3,4, 6$ and the total dimension
$D = 11, 10, 11$, respectively. (We thus obtain as well one exact solution
for ten-dimensional superstring gravity \cite{GSW} and two solutions for
eleven-dimensional supergravity \cite{Cr} and $M$-theory \cite{LPX}.)
These solutions are generalized
to arbitrary signatures of the $N_0$-dimensional section
of the metric. The Riemann tensor squared for the solutions
is calculated and it is proved that for non-Euclidean signatures
it is divergent  on a certain (generalized) hypersphere
in   ${\bf R}^{N_0}$. An example  of a de-Sitter membrane solution
is suggested. In \sect 5  we consider the exceptional case
$N_0 =2$. We show that in this case the midisuperspace metric is not
uniquely determined and depends on the choice of the conformal frame.
(As pointed out in \Ref{RZ}, there is no conformal Einstein-Pauli
frame in this case). Two examples corresponding
to different conformal frames are presented.

\psection{The model}

Let us consider the manifold
\beq             %2.1
M = M_{0}  \times M_{1} \times \ldots \times M_{n},
\eeq
with the metric
\beq          %2.2
g= \e^{2{\gamma}(x)} g^0  +
\sum_{i=1}^{n} \e^{2\phi^i(x)} g^i ,
\eeq
where
\beq          %2.3
g^0  = g^0 _{\mu \nu}(x) dx^{\mu} \otimes dx^{\nu}
\eeq
is a metric on the manifold $M_{0}$ and $g^i $ is a metric on
$M_{i}$  satisfying the equation
\beq           %2.4
{R_{m_{i}n_{i}}}[g^i ] = \lambda_{i} g^i _{m_{i}n_{i}},
\eeq
$m_{i},n_{i}=1,\ldots, N_{i}$; $\lambda_{i}= {\rm const}$,
$i=1,\ldots,n$.  Thus $(M_i, g^i )$  are Einstein spaces.
The functions $\gamma, \phi^{i} : M_0 \rightarrow {\bf R}$ are smooth.

\medskip\noi
{\bf Remark 1.} It is more correct to write (2.2) as
\[
g= \exp[2{\gamma}(x)] \hat g^0  +
     \sum_{i=1}^{n} \exp[2{\phi^{i}}(x)] \hat g^i
\]
where  we denote
by $\hat{g}^{\alpha} =  p_{\alpha}^{*} g^{\alpha}$ the
pullback of the metric $g^{\alpha}$  to the manifold  $M$
by the canonical projection: $p_{\alpha} : M \rightarrow M_{\alpha}$,
$\alpha = 0, \ldots, n$. In what follows
all "hats" over metrics will be omitted.

Here we  are interested in exact solutions to the Einstein
equations with a cosmological constant
\beq      %2.5
R_{MN}[g] - \half g_{MN} R[g]  =  - \Lambda g_{MN}
\eeq
for the metric (2.2) defined on the manifold (2.1).
The set of equations (2.5) is equivalent to
\be %2.6
R_{MN}[g]  =  \frac{2 \Lambda}{D-2} g_{MN},
\ee
where  $D = \sum_{k=0}^{n} N_k = \dim  M$
is the dimension of the manifold (2.1), $N_{k} = \dim  M_{k}$,
$k=0,\ldots,n$. \eqs (2.5) are the field equations corresponding to the action
\bear           %2.7
S\lal  =  S[g]        \nn
\lal = \frac{1}{2 \kappa^{2}}
     \int_{M} d^{D}x \sq g \{ R[g] -  2 \Lambda \}   + S_{\rm GH}
\ear
where we denote $|g| = |\det (g_{MN})|$:
$S_{\rm GH}$ is the standard Gibbons-Hawking boundary term
\cite{GH}. This term is essential for a quantum treatment
of the problem.

The nonzero Ricci tensor components
for the metric (2.2) are  (see the Appendix)
\bearr            %2.8 -2.9
R_{\mu \nu}[g]  =   R_{\mu \nu}[g^0 ] +
          g^0 _{\mu \nu} \Bigl[- \Delta_0 \gamma
          +(2-N_0)  (\p \gamma)^2                    \nnn \quad
- \p \gamma \sum_{j=1}^{n} N_j \p \phi^j ]
+ (2 - N_0) (\gamma_{;\mu \nu} - \gamma_{,\mu} \gamma_{,\nu}) \nnn \quad
 - \sum_{i=1}^{n} N_i ( \phi^i_{;\mu \nu} - \phi^i_{,\mu} \gamma_{,\nu}
 - \phi^i_{,\nu} \gamma_{,\mu} + \phi^i_{,\mu} \phi^i_{,\nu}),  \\ \lal
R_{m_{i} n_{i}}[g]  = {R_{m_{i} n_{i}}}[g^i ]
     - \e^{2 \phi^{i} - 2 \gamma} g^i _{m_{i} n_{i}}
      \biggl\{ \Delta_0 \phi^{i} \nnn \quad
+ (\p \phi^{i}) [ (N_0 - 2) \p \gamma  +
          \sum_{j=1}^{n} N_j \p \phi^j ] \biggr\},
\ear
Here
$\p \beta \,\p \gamma \equiv g^{0\  \mu \nu} \beta_{, \mu} \gamma_{, \nu}$
and  $\Delta_0$ is the Laplace-Beltrami operator corresponding
to  $g^0 $. The scalar curvature for (2.2) is
\bearr            %2.10
  \nq R[g] =  \sum_{i =1}^{n} \e^{-2 \phi^i} {R}[g^i ]
          + \e^{-2 \gamma} \biggl\{ {R}[g^0 ]
          - \sum_{i =1}^{n} N_i (\p \phi^i)^2  \nnn
 -  (N_0 {-} 2) (\p \gamma)^2
     -  (\p f)^2 - 2 \Delta_0 (f +  \gamma) \biggr\}
\ear
where
\beq  %2.11
f = {f}(\gamma, \phi)  = (N_0 - 2) \gamma +\sum_{j=1}^{n} N_j  \phi^j .
\eeq

Using (2.8) and (2.9), it is not difficult to verify that the
field equations (2.5) (or, equivalently, (2.6)) may be obtained
as the equations of motion corresponding to the action
\bear           %2.12
\lal S_{\sigma}[g^0 , \gamma,\phi]      \nn
\lal =   \frac{1}{2 \kappa^{2}_0}
     \int_{M_0} d^{N_0}x   \sq {g^{0}}
                                     %%|{\rm det} (g^0 _{..})|^{1/2}
     \e^{f(\gamma, \phi)} \biggl\{ {R}[g^0 ] \nnn
- \sum_{i =1}^{n} N_i (\p \phi^i)^2  -  (N_0 - 2) (\p \gamma)^2
           + (\p f) \p (f + 2\gamma) \nnn
+      \sum_{i=1}^{n} \lambda_{i} N_i \e^{-2 \phi^i + 2 \gamma} -
     2 \Lambda \e^{2 \gamma} \biggr\}.
\ear
where $|g^{0}|= |\det (g^0 _{\mu\nu})|$ and similar notations are applied
to the metrics $g^{i}$, $i=1, \ldots, n$.
For finite internal space volumes (e.g. compact $M_i$)
\beq   %2.13
V_i = \int_{M_i} d^{N_i}y \sq{g^i } < + \infty,
\eeq
the action (2.12) coincides with the action (2.7), i.e.
\beq   %2.14
     S_{\sigma}[g^0 , \gamma,\phi] =  S[g],
\eeq
where $g$ is defined by the relation (2.2) and
\beq    %2.15
\kappa^{2} = \kappa^{2}_0 \prod_{i=1}^{n} V_i.
\eeq
This may be readily verified using the following relation for
the scalar curvature (2.10):
\bearr            %2.14
\nq R[g] {=}  \sum_{i=1}^{n} \! \e^{-2 \phi^i} {R}[g^i ]
          + \e^{-2 \gamma} \Bigl\{ {R}[g^0 ]
          - \sum_{i =1}^{n} N_i (\p \phi^i)^2  \nnn
- (N_0 - 2) (\p \gamma)^2 + (\p f) \p (f + 2 \gamma) + R_{B} \Bigr\},
\ear
where
\bearr %2.17
R_B = (1/\sq {g^0 }) \e^{-f}
     \p_{\mu} [-2 \e^f \sq {g^0 }
     g^{0\ \mu \nu} \p_{\nu} (f + \gamma)]  \nnn
\ear
gives rise to the Gibbons-Hawking boundary term
\beq %2.18
S_{\rm GH} = \frac{1}{2\kappa^{2}} \int_{M} d^{D}x \sq g
     \{ - \e^{-2 \gamma} R_{B} \}.
\eeq

\psection{The non-exceptional case $N_0 \neq 2$}

In order to simplify the action (2.12), we use,
as in \cite{Ber}  for $N_0 \neq 2$, the gauge
\beq %3.1
\gamma = {\gamma}_{0}(\phi) =
\frac{1}{2-N_0}  \sum_{i =1}^{n} N_i \phi^i.
\eeq
It means that $f = {f}(\gamma_0, \phi)= 0$, or the conformal
Einstein-Pauli frame is used.
Evidently this frame does not exist for $N_0 = 2$.
For the cosmological case  $N_0 =1$,  $g^0  = - dt \otimes dt$,
and (3.1) corresponds to the harmonic-time gauge  \cite{IMZ}.
>From (3.1) we get
\bearr           %3.2
S_{0}[g^0 ,\phi] = S_{\sigma}[g^0 , \gamma_0,\phi] =  \nnn
\frac{1}{2 \kappa^{2}_0}
     \int_{M_0} d^{N_0}x \sq {g^0 } \Bigl\{ {R}[g^0 ] \nnn
- G_{ij} g^{0\ \mu \nu} \p_{\mu} \phi^i  \p_{\nu} \phi^j -  2 {V}(\phi) \},
\ear
where
\beq %3.3
G_{ij} = N_i \delta_{ij} + \frac{N_i N_j}{N_0 -2}
\eeq
are the components of the ``midisuperspace" (or target space) metric on
${\bf R}^{n}$
\beq %3.4
G = G_{ij} d \phi^{i} \otimes d \phi^{j}
\eeq
and
\beq
\nq V = {V}(\phi) = \Lambda \e^{2 {\gamma_0}(\phi)} - \half
     \sum_{i =1}^{n} \lambda_i N_i \e^{-2 \phi^i + 2 {\gamma_0}(\phi)}
\eeq
is the potential. (Here we corrected a misprint in \eq (11) from
\cite{Ber}.) Thus, we are led to the action of a self-gravitating
$\sigma$ model with a flat target space
$({\bf R}^{n}, G)$ (3.4) and a self-interaction described by the
potential (3.5).

For $N_0 =1$,  $g^0  = - dt \otimes dt$ the action (3.2) coincides with
the well-known cosmological one  \cite{IMZ}. In this case the
minisuperspace metric (3.3) is pseudo-Euclidean \cite{IM1,IMZ}.

\medskip\noi
{\bf Remark 2.}
We note that in the infinite-dimensional case $n = \infty$
the potential (3.5) is well-defined if the following restrictions
are imposed:
\beq  %3.6
\sum_{i =1}^{n} |\lambda_{i}| N_i < + \infty, \qquad
                \sum_{i =1}^{n}  N_i |\phi^i|< + \infty.
\eeq
In the case $N_0 =1$, \ $\phi = (\phi^i)$ belongs to a Banach space with
$l_1$-norm \cite{I}.

\psubsection{The case $N_0 > 2$}

For  $N_0 > 2$  the midisuperspace metric (3.3) is Euclidean.
The potential (3.5) may be rewritten as
\be %3.7
{V}(\phi) =
\sum_{\alpha = 0}^{n} A_{\alpha} \exp[u_i^\alpha \phi^i],
\ee
including the cosmological constant and the curvature terms,
where $A_0 = \Lambda$, $A_{j}= - \frac{1}{2}  \lambda_{j} N_j$ and
\beq %3.8
u_i^0  = \frac{2N_i}{2- N_0}, \qquad
u_i^j  = 2 \Bigl( - \delta^j_i + \frac{N_i}{2- N_0}\Bigr),
\eeq
$i, j = 1, \ldots, n$. Thus the potential (3.5) has a Toda-like form.

Let
\beq %3.9
\avers{u,v} \ \equiv G^{ij} u_i v_j
\eeq
be a quadratic form on  ${\bf R}^n$. Here
\beq %3.10
G^{ij} =
\frac{\delta_{ij}}{N_i} + \frac{1}{2 - D}
\eeq
are components of the matrix inverse to the matrix  $(G_{ij})$ in
(3.3). For the vectors (3.8) $u^\alpha = (u^\alpha_i) \in {\bf R}^n$,
$\alpha = 0, \ldots, n,$ we get the following relations:
\bear           %3.11 -13
\lal \avers{u^0 , u^0 } = \frac{4(D - N_0)}{(N_0 - 2)(D -2)}, \\
\lal \avers{u^0 , u^j } = \frac{4}{(N_0 - 2)},                 \\
\lal \avers{u^i , u^j } =
          4\biggl(\frac{\delta_{ij}}{N_i} + \frac{1}{N_0 - 2}\biggr),
\ear
$i, j = 1, \ldots, n$.

\psubsection{The Adler-van-Moerbeke criterion}

For a fixed metric $g^0 $ the action (3.2) coincides with the
action of a Euclidean Toda-like system, i.e. a dynamical (physical)
system with the potential in the form of a sum of exponents
depending on linear combinations of cordinates (fields).
For Toda-like systems in the dimension $N_0 = 1$
\cite{B}-\cite{OP} (with the appropriate number of exponents)
we know that the integrable
cases (open and closed Toda lattices)
occur when the vectors  $u^\alpha$  in the exponents
correspond to roots of an appropriate finite-dimensional
semisimple Lie algebra or an infinite-dimensional affine Lie algebra.

This situation may be described by the
so-called Adler-van-Moerbeke criterion  \cite{AM}. Here
we formally extend this criterion to the case $N_0 > 1$ and
apply it to our model with a fixed metric $g^0 $.

When all $A_{\alpha} \neq 0$ in
(3.5) and the vectors  $u^\alpha$  satisfy the
Adler-van-Moerbeke criterion  \cite{AM},
\beq            %3.14
K_{\alpha \beta} \equiv
\frac{2\avers{u^\alpha, u^\beta}}
     {\avers{u^\beta, u^\beta}} = \hat{C}_{\alpha \beta},
\ee
$\alpha = 0, \ldots, n$, where $\hat{C} =  (\hat{C}_{\alpha \beta})$ is
the Cartan matrix corresponding to some
affine Lie algebra  $\hat{{\cal G}}$
\cite{KM}, then the considered Toda-like system (3.2) with fixed
$g^0 $ is  equivalent  to an  $N_0$-dimensional
closed Toda lattice on $(M_0, g^0 )$ corresponding to $\hat{{\cal G}}$.

When $\Lambda = 0$, $\lambda_i \neq 0$, $i = 1, \ldots, n$, $n \geq 2$ and
\beq            %3.15
K_{ij} = C_{ij},
\ee
$i, j = 1, \ldots, n$, where  $C = (C_{ij})$  is  the Cartan matrix
corresponding to some semisimple Lie algebra  ${\cal G}$  of rank $n$,
then the Toda-like system (3.2) with fixed $g^0 $ is equivalent to an
$N_0$-dimensional open Toda lattice on  $(M_0, g^0 )$
corresponding to  ${\cal G}$.

Now, we show that the relations (3.14) and (3.15) are not
satisfied for $N_i \in {\bf N}$ ($N_i > 1$,
since $\lambda_i \neq 0$), $i = 1, \ldots, n$, $n \geq 2$.
Indeed, from (3.13) we get
\beq            %3.16
K_{ij} = \frac{2\bigl[\delta_{ij}/N_j + 1/(N_0 - 2)\bigr]}
     {1/N_j + 1/(N_0 - 2)} > 0,
\eeq
It follows from (3.16) that the relation (3.15)
is never satisfied for $N_i \in {\bf N}$, since
\beq            %3.17
C_{ij} = - n_{ij}, \qquad  n_{ij} \in {\bf Z}_{+} = \{0,1,2, \ldots \},
\ee
for $i \neq j$ ($n_{ij} = 0,1,2,3$).
For the same reason ($\hat{C}_{\alpha \beta} \leq 0, \alpha \neq \beta$)
the relation (3.14) is never satisfied for positive integers
$N_j$ and  $n \geq 1$ (see (3.12)).  Thus, the model under consideration
(3.2) (with fixed $g^0 $) is not equivalent to an $N_0$-dimensional
(closed or open) Toda lattice (when the number of nonzero terms in
the potential (3.5) is greater than one) and seems to be a rather nontrivial
object of non-linear analysis.

\medskip\noi
{\bf Remark 3.} If we consider (at least formally)
the model (3.2) with $\Lambda = 0$ and complex dimensions
$N_j$, $j = 1, \ldots, n$, obeying the restriction
\beq            %3.18
  \det (G_{ij}) = N_1 \ldots N_n \frac{2 - D}{2 - N_0} \neq 0,
\ee
then we find the following solution of (3.15): $n = 2$,
\beq            %3.19
\{ N_1, N_2 \} = \biggl\{ \frac{1}{3}(2-N_0), \frac{k}{k+2} (2- N_0)
    \biggr\}
\ee
$k = 1, 2, 3$, corresponding to the Lie algebras   $a_2 = sl(3)$,
$b_2 = so(5)$ and $g_2$, respectively. (The cosmological case $N_0 =1$ was
considered earlier in \Ref{I}. For $N_0 =1$, $k =1$  see also \Ref{GM}.)

\psubsection{ Diagonalization}

{\bf The case $N_0 > 2$.} Let us diagonalize the midisupermetric.
This may be useful for quantization of the $\sigma$ model under study.
For  $N_0 > 2$  the midisupermetric
may be diagonalized by the linear transformation
\beq            %3.20
\varphi^a = S^a_i \phi^i,
\ee
where
\beq            %3.21
S^a_i  \delta_{ab}  S^b_j = G_{ij},
\ee
$a, b = 1, \ldots, n$; $i, j = 1, \ldots, n$. Then \eq (3.4) reads:
\beq            %3.22
G = \delta_{ab} d \varphi^{a} \otimes d \varphi^{b} .
\ee

An example of diagonalization (3.20), (3.21) is
\bear           %3.23 -24
\lal \varphi^1 = q^{-1} \sum_{i=1}^{n} N_i \phi^i, \\
\lal \varphi^{\hat{b}} =
\Big[N_{\hat{b}-1}/(\Sigma_{\hat{b}-1} \Sigma_{\hat{b}})  \Big]^{1/2}
\sum_{j= \hat{b}}^{n} N_j (\phi^j {-} \phi^{ \hat{b}-1})   ,
\ea
$\hat{b} = 2, \ldots,n$, where
\bearr             %3.25
\nq q = q(N_0,D) \equiv \Big[\frac{(D{-}N_0)|N_0{-}2|}{(D -2)} \Big]^{1/2},
\ \Sigma_a \equiv  \sum_{j=a}^{n} N_j.   \nnn
\ear

Consider a more general class of the diagonalization (3.20)
satisfying (3.23) or, equivalently,
\beq             %3.26
S^1_i = q^{-1} N_i,
\ee
Let us introduce
\beq             %3.27
S^a = (S^a_i)  \in {\bf R}^{n},
\ee
$a = 1, \ldots, n$. The relation (3.21) is equivalent to
\beq             %3.28
S^a_i G^{ij} S^b_j =  \avers{S^a, S^b} = \delta^{ab}.
\ee

For $a, b = 1$ the relation (3.28) is satisfied identically due to
(3.25) and (3.26) (see also (3.8), (3.11)). For $\hat{b} > 1$
\bearr             %3.29
0 = \avers{S^1, S^{\hat{b}}} = q^{-1} N_i G^{ij} S^{\hat{b}}_j
     =   q^{-1} \frac{2 - N_0}{2 -D} \sum_{j=1}^{n} S^{\hat{b}}_j,\nnn
\ear
or, equivalently,
\beq             %3.30
0 =  \sum_{j=1}^{n} S^{\hat{b}}_j.
\ee
Here we use the relation
\beq             %3.31
G^{ij} N_j  =  \frac{2 - N_0}{2 -D}.
\ee

For $\hat{a}, \hat{b} > 1$  we get from (3.30)
\bearr            %3.32
\delta^{\hat{a} \hat{b}}  = \avers{S^{\hat{a}}, S^{\hat{b}}}
= \Big( \frac{\delta_{ij}}{N_i} + \frac{1}{2 - D} \Big)
S^{\hat{a}}_i S^{\hat{b}}_j = \frac{\delta_{ij}}{N_i}
S^{\hat{a}}_i S^{\hat{b}}_j  \nnn
\ear
or, equivalently,
\beq             %3.33
\sum_{i=1}^{n}  \frac{1}{N_i} S^{\hat{a}}_i S^{\hat{b}}_i
= \delta^{\hat{a} \hat{b}}.
\ee

Thus, when the condition (3.26) is imposed, the relation (3.21)
is equivalent to the set of relations (3.30), (3.33). It is
not difficult to verify that these relations are satisfied for
$(S^{\hat{a}}_i)$ from (3.24).
For the inverse matrix we get from (3.28)
\beq             %3.34
\hat{S}^i_a =  G^{ij} S^{b}_j \delta_{ba} = G^{ij} S^{a}_j
\ee
and, hence, (see (3.26) and (3.31))
\beq             %3.35
\hat{S}^i_1 =  G^{ij} S^{1}_j  =  q^{-1} \frac{2 - N_0}{2 -D}
=  \frac{q}{D-N_0}.
\ee

>From the relation
\beq             %3.36
\hat{S}^i_a G_{ij} \hat{S}_b^j =  \delta_{ab}
\ee
(following from (3.28)) and \eqs (3.10), (3.35), (3.36) we get
\beq             %3.37
\sum_{j=1}^{n}  N_j \hat{S}^{j}_{\hat{b}} = 0, \qquad
\sum_{i=1}^{n}  N_i \hat{S}^{i}_{\hat{a}} \hat{S}^{i}_{\hat{b}}
= \delta_{\hat{a} \hat{b}},
\ee
$\hat{a}, \hat{b} > 1$. Here we have used the relation
\beq             %3.38
\sum_{i=1}^{n}  G_{ij} = N_j \frac{D - 2}{N_0 - 2}.
\ee

In the new variables (3.20) satisfying (3.26) the action (3.2) reads:
\bearr %3.39
S =  \frac{1}{2 \kappa^{2}_0}
\int_{M_0} d^{N_0}x \sq{g^0 } \{ {R}[g^0 ] \nnn
\qquad - \delta_{ab} g^{0\ \mu \nu} \p_{\mu} \varphi^a  \p_{\nu} \varphi^b
-  2 V \}.
\ear
where
\beq            %3.40
V =\sum_{\alpha = 0}^{n} A_{\alpha} \exp[\hat{u}_a^\alpha \varphi^a].
\ee
Here the following notation is used:
\beq            %3.41
\hat{u}_a = S_a^i u_i.
\ee
It follows from (3.35) that
\beq            %3.42
\hat{u}_1 = \hat{S}_1^i u_i =  \frac{q}{D-N_0} \sum_{i=1}^{n} u_i.
\ee
For the vectors  (3.8), corresponding to
the $\Lambda$-term and the curvature components, respectively, we have
\beq            %3.43
\hat{u}^0 _1 = \frac{2q}{2-N_0}, \qquad
\hat{u}^j _1 = - 2q^{-1},
\ee
$j = 1, \ldots, n$.
We denote $\vec{u}_{*} = (\hat{u}_2, \ldots, \hat{u}_n)$.
Then $\vec{u}^0 _{*} = 0$  (see (3.37)) and
\bearr %3.44
\vec{u}^i _{*} \vec{u}^j _{*}=\avers{u^i , u^j } + 4 q^{-2}\nnn
\cm\cm = 4 \Big( \frac{\delta_{ij}}{N_i} + \frac{1}{N_0 - D} \Big),
\ear
$i, j = 1, \ldots, n$ (see (3.13), (3.43)).
Thus the potential (3.40) (see (3.5)) may be written as
\beq            %3.45
V =  \Lambda \exp \Big[ \frac{2q \varphi^{1}}{2-N_0} \Big] +
\exp( - 2q^{-1} \varphi^{1} ) {V_{*}}(\vec{\varphi}_{*})
\ee
where
\beq            %3.46
{V_{*}}(\vec{\varphi}_{*}) = \sum_{i =1}^{n} ( - \frac{1}{2} \lambda_{i} N_i)
\exp(\vec{u}^i _{*} \vec{\varphi}_{*}),
\ee
$\vec{\varphi}_{*} = (\varphi_{2}, \ldots, \varphi_{n})$
and the vectors  $\vec{u}^i _{*} \in {\bf R}^{n-1}$ satisfy the relations
(3.44).

\medskip\noi
{\bf The cosmological case $N_0 =1$.}  In the cosmological case
$M_0 = {\bf R}$, $g^0  = - {\cal N}^2(t) dt \otimes dt$,
(${\cal N}(t) > 0$ is the lapse function) for the metric (2.2)
\beq          %3.47
g= - \e^{2\gamma_0 (t)} {\cal N}^2(t) dt \otimes dt +
\sum_{i=1}^{n} \e^{2\phi^i (x)} g^i
\eeq
the action (3.2) reads \cite{IMZ}:
\bearr           %3.48
S = S[{\cal N},\phi] =
\frac{1}{\kappa^{2}_0} \int dt
{\cal N} \{ \half {\cal N}^{-2} \bar{G}_{ij}
\dot{\phi}^{i} \dot{\phi}^{j} -   {V}(\phi) \}, \nnn
\ear
where
\beq            %3.49
\bar{G}_{ij} = N_i \delta_{ij} - N_i N_j
\ee
are components of a pseudo-Euclidean minisuperspace  metric on
${\bf R}^{n}$ and the potential $V$ is defined in (3.5).

Let us consider the diagonalization
\beq            %3.50
\varphi^a = S^a_i \phi^i, \qquad
S^a_i  \eta_{ab}  S^b_j = G_{ij},
\ee
($(\eta^{ab}) = {\rm diag}(-1, 1, \ldots, 1$),
$a, b = 0, \ldots, n-1$; $i, j = 1, \ldots, n$)
satisfying \eq (3.26) with $q$ from (3.25)
($N_0 =1$). Just as before, it may be shown that
in the new variables $\varphi^a $ the action (3.48) has the form
\beq           %3.51
\nq S = S[{\cal N},\phi] =  \frac{1}{\kappa^{2}_0} \int dt
{\cal N} \{\half {\cal N}^{-2}
\eta_{ab} \dot{\varphi}^{a} \dot{\varphi}^{b} -   V \}
\ee
with the potential (3.5) rewritten in the new variables
\beq            %3.52
V = \Lambda \exp [ 2q \varphi^{0}] +
        \exp(2q^{-1} \varphi^{0} ) {V_{*}}(\vec{\varphi}_{*}),
\ee
where ${V_{*}}(\vec{\varphi}_{*})$ is defined in (3.46),
the vectors  $\vec{u}^i _{*} \in {\bf R}^{n-1}$ satisfy
the relations (3.44) with $N_0 = 1$, and $\vec{\varphi}_{*} =
(\varphi_{1}, \ldots, \varphi_{n-1})$.

\psection{Exact solutions}

Here we consider the metric (2.2) defined on the
manifold (2.1) with the relations (2.4) and
\beq          %4.1
M_0 = {\bf R}^{N_0}, \qquad
g^0  = \sum_{a=1}^{N_0} dx^{a} \otimes dx^{a},
\eeq
assuming $N_0 > 2$. Thus the $N_{0}$-dimenional section
of the metric (2.2) is conformally flat.
One of the simplest Ans\"atze for (2.2) is the following:
\beq            %4.2
\gamma = \alpha_0 {u}(|x|^2), \qquad
\phi^i = \alpha_i {u}(|x|^2) + \beta_i,
\ee
where  $\alpha_0, \alpha_{i}, \beta_i$ are constants,
$i = 1, \ldots, n$, and $|x|^2 = \sum_{a=1}^{N_0} (x^{a})^2$.
We are interested in spherically symmetric solutions to the Einstein
equations (2.5) with $\Lambda = 0$ governed by the function $u = {u}(z)$
and the parameters $\alpha_{\nu}, \beta_i$. The field equations
\beq            %4.3
R_{MN}[g] = 0
\ee
for the metric (2.2) satisfying (4.1) and (4.2), are
equivalent to the following set of equations:
\bearr           %4.4 - 4.6
A \equiv - \alpha_0 (4z u'' + 2 N_0 u') \nnn
\cm \cm +
4 \alpha_0 \hat{\alpha} z (u')^2 + 2 \hat{\alpha} u' = 0, \\
     \lal B \equiv \hat{\alpha} u'' + [ (N_0 -2) \alpha^2_0 \nnn
\qquad +
     2 \alpha_0  \sum_{j =1}^{n} N_j \alpha_j
- \sum_{j =1}^{n} N_j \alpha_j^2 ] (u')^2 = 0, \\
\lal C_i \equiv \lambda_i
- \alpha_i \e^{2(\alpha_i - \alpha_0) u + 2 \beta_i} \times \nnn
\cm  \times [4 z u'' + 2 N_0 u' -  4 \hat{\alpha} z (u')^2] = 0,
\ear
$i = 1, \ldots, n$. Here $u' = {du}/{dz}$, $u'' = {d^2 u}/{dz^2}$
and
\beq            %4.7
\hat{\alpha} =  (2 -N_0) \alpha_0- \sum_{j =1}^{n} N_j \alpha_j.
\ee

The reduction of (4.3) to \eqs (4.4)-(4.6)
takes place due to the following representation for the Ricci
tensor components (2.8) and (2.9) in our case (4.2):
\bearr           %4.8 -4.9
R_{ab}[g] = A \delta_{ab} + 4B x^a x^b, \\
\lal R_{m_i n_i}[g] = C_i g^i _{m_i n_i},
\ea
$a,b = 1, \dots, N_0$;  $i = 1, \dots, n$.

Here we adopt the following Ansatz for the function $u(z)$ from (4.2):
\beq            %4.10
u(z) = \ln(C + z),
\ee
where $C$ is a constant. Under the substitution (4.10) \eq (4.4)
is satisfied identically if
\beq            %4.11
\hat{\alpha} = -1, \qquad \alpha_0 = -1/N_0.
\ee
(We note that $u'' = - (u')^2$. For  $C \neq 0$, (4.4) implies (4.11).)
Then, (4.4) and (4.5) read:
\bearr           %4.12 - 4.13
\sum_{j =1}^{n} N_j \alpha_j = 2 -  \frac{2}{N_0}, \\
\lal \sum_{j =1}^{n} N_j \alpha_j^2 = \frac{(N_0 -1)(N_0 -2)}{N_0^2}.
\ear
\eqs (4.6) are equivalent to the relations
\beq            %4.14
2(\alpha_0 - \alpha_i) = -1, \qquad
2 N_0 \alpha_i \e^{2 \beta_i}  = \lambda_i,
\ee
$i =1, \ldots, n $. From (4.11) and (4.14)  we obtain
\beq            %4.15
\alpha_i = \frac{1}{2} -  \frac{1}{N_0}, \qquad
\e^{2 \beta_i} = \frac{\lambda_i}{N_0 - 2} \neq 0.
\ee
A substitution of (4.15) into (4.12),
(4.13) gives the following Diophantus equation
for the dimensions $N_{\nu}$:
\beq            %4.16
\sum_{j =1}^{n} N_j  =  \frac{4(N_0 - 1)}{N_0 -2}.
\ee
\eq (4.16) has the solutions
\beq            %4.17
\sum_{j =1}^{n} N_j  =  8, 6, 5 \quad {\rm for} \quad  N_0 = 3, 4, 6,
\ee
respectively.
>From (2.2), (4.1), (4.2), (4.10), (4.11) and (4.15) we obtain the metric
\bearr           %4.18
 g = \bigl[C {+} |x|^2\bigr]^{1 - 2/N_0}
\biggl[ \sum_{a=1}^{N_0} \frac{dx^{a} {\otimes} dx^{a}}{C + |x|^2}
     + \sum_{i=1}^{n} \frac{\lambda_i}{N_0 {-} 2} g^i  \biggr]\nnn
\ear
defined on the manifold
\beq           %4.19
M = {\bf R}_C^{N_0} \times M_{1} \times \ldots \times M_{n},
\eeq
where
\beq            %4.20
{\bf R}_C^{N_0} = \{x \in {\bf R}^{N_0} : C + |x|^2 > 0 \}
\subset {\bf R}^{N_0}
\ee
is an open domain  in ${\bf R}^{N_0}$,   $C \in {\bf R}$.
The metric (4.18) describes, for $N_0 = 3, 4, 6$, three
families of spherically symmetric ($O(N_0)$-symmetric)
solutions to the vacuum Einstein equations  (4.3)
with $n$ internal Einstein spaces of nonzero  curvature $(M_i, g^i )$
(2.4).  It follows from (4.16), (4.17) that
\bear           %4.21 -22
\nhq    D \eql N_0 + \sum_{j =1}^{n} N_j =
                       \frac{N_0^2}{N_0 - 2} + 2 =11, 10, 11,\\
                 \lal \cm n \leq n_0 = 4, 3, 2
\ear
for $N_0 = 3, 4, 6$, respectively.

\psubsection{ Nonsingular solutions}

For $C > 0$, ${\bf R}_C^{N_0}
= {\bf R}^{N_0}$  and the metric (4.18) describes spherically symmetric
nonsingular solutions to the Einstein equations
defined on the manifold
\beq           %4.23
{\bf R}^{N_0} \times M_{1} \times \ldots \times M_{n}.
\eeq
(It should be stressed that
the $N_0$-dimensional part of the metric (4.18) has Euclidean signature.)
A special case of this solution with $N_0 = 6$, $n = 1$, $N_1 = 5$ was
recently considered in \cite{GIKM}.

\psubsection{Exceptional solutions}

Let us consider the solution (4.18)
with $C = 0$. It can be written as follows:
\beq            %4.24
g = d \rho \otimes d \rho  +   \rho^2 g_{*},   \qquad
         \rho = \alpha^{-1} |x|^{\alpha}
\eeq
where   $ \alpha = 1 - 2/N_0 $ and
\beq            %4.25
g_{*} = \alpha^2  \biggl[ g(S^{N_0 - 1})
     + \sum_{i=1}^{n} \frac{\lambda_i}{N_0 - 2} g^i  \biggr]
\ee
is the Einstein metric on the manifold
\beq            %4.26
M_{*} =
S^{N_0 - 1} \times M_{1} \times \ldots \times M_{n}.
\ee
Here  ${g}(S^{N_0 - 1})$ is the canonical metric on an
$(N_0 {-} 1)$-dimensional sphere $S^{N_0 - 1}$.
The metric $g_{*}$ in (4.24) satisfies the relation
\beq            %4.27
{\rm Ric}\, [g_{*}] = (D - 2) g_{*},
\ee
where   ${\rm Ric}\ [g_{*}]$ is the Ricci tensor corresponding to
$g_{*}$  and $D= \dim M$.
The metric (4.24) is defined on the manifold ${\bf R}_{+} \times M_{*}$
(see Remark 1) and is non-flat, as may be verified using
the relations (6.2)-(6.4) from the Appendix.
The $N_0$-dimensional section of the metric is also non-flat
(due to "deficit" of the spherical angle).
Since the solution (4.24) is an attractor for (4.18) as
$|x| \to \infty$, we see that the metric (4.18) and
its $N_0$-dimensional section  have non-flat asymptotics.

\psubsection{Solutions with arbitrary signature}

The solution (4.18) may be considered as a special case of the
following solutions with arbitrary signature of ``our" space:
\bearr           %4.28
g = \bigl[C + \eta_{ab} x^a x^b\bigr]^{1 - 2/N_0}
\biggl\{ \frac{\eta_{ab}dx^{a} \otimes dx^{b}}{C + \eta_{ab} x^a x^b} \nnn
\cm \cm  + \sum_{i=1}^{n} \frac{\lambda_i}{N_0 - 2} g^i  \biggr\}.
\ear
Here
\beq            %4.29
\eta = (\eta_{ab}) = {\rm diag} (w_1, \ldots, w_{N_0}),\qquad w_a = \pm 1.
\ee
The metric (4.28) is defined on the manifold
\beq           %4.30
M = {\bf R}_{C, \eta}^{N_0} \times M_{1} \times \ldots \times M_{n},
\eeq
where
\beq            %4.31
{\bf R}_{C,\eta}^{N_0} =
\{x \in {\bf R}^{N_0} : C + \eta_{ab} x^a x^b > 0 \} \subset {\bf R}^{N_0}
\ee
is supposed to be non-empty (i.e the case when $C < 0$ and all
$w_a = -1$ in (4.29) is excluded). The metric (4.28) satisfies
the vacuum Einstein equations (4.3). It may be obtained from
(4.18) by a Wick-type rotation, i.e. we write $x^a =
w_a^{1/2} \hat{x}^a$, $w_a > 0$,
in (4.18) and then perform an analytical continuation in $w_a$.

\medskip\noi
{\bf Proposition 1}.  The Riemann tensor squared for the metric
(4.28) has the form
\bearr           %4.32
{I}[g] \equiv {R_{MNPQ}}[g] {R^{MNPQ}}[g] \nnn
\cm     = (C + x^2)^{-2 -2 \alpha} (\bar{I}_1  + \bar{I}_2),
\ear
where
\bearr           %4.33 - .34
\nq \bar{I}_1 = (\alpha -1)^2 (N_0 -1) \{ 16 C^2 \nnn
    \cm  + 2 (N_0 - 2) [2C + (\alpha +1) x^2]^2 \},    \yy
\lal \nq \bar{I}_2 =  - 4 \alpha^2 N (N_0 - 2) x^2 (C + x^2)  \nnn
  + (C + x^2)^2 \sum_{i = 1}^{n}
\Bigl (\frac{N_0 {-}2}{\lambda_i}\Bigr )^2 I [g^i ]
     +  2 \alpha^4 N (N {-} 1) (x^2)^2                 \nnn
    \cm  + 4 \alpha^2 N (N_0 - 1)      (\alpha x^2 + C)^2;
\ear
here $\alpha = 1 - 2/N_0$, $x^2 = \eta_{ab} x^a x^b$,
$N = \sum_{j =1}^{n} N_j$  and
${I}[g^i ]$ is the Riemann tensor squared for the metric $g^i $.

\medskip\noi
{\bf Proof.}  \eqs (4.32)-(4.34) may be obtained
using the formula (6.10) from the Appendix. But a simpler way
is to calculate first the Riemann tensor squared in the Euclidean case
$\eta_{ab} = \delta_{ab}$,
\bearr           %4.35
g = [C + r^2]^\alpha
\biggl\{ \frac{dr \otimes dr + r^2 d \Omega^2_{N_0 -1}}{C + r^2}   \nnn
 \cm\cm  + \sum_{i=1}^{n} \frac{\lambda_i}{N_0 - 2}\ g^i  \biggr\}
\ear
where $r^2 = \delta_{ab} x^a x^b$ and $d \Omega^2_{N_0 -1}
= {g}(S^{N_0 - 1})$ is
the metric on $S^{N_0 -1}$, using the ``cosmological" relation  (6.15)
from the Appendix,  and then perform the Wick rotation
$r^2 \to \eta_{ab} x^a x^b$.

\medskip\noi
{\bf Proposition 2.} For the metric (4.28) with
a non-Euclidean signature $(\eta_{ab}) \neq (\delta_{ab})$ and $C \neq 0$
\beq            %4.36
{R_{MNPQ}}[g] {R^{MNPQ}}[g] \to + \infty
\ee
as  $C + \eta_{ab} x^a x^b \to + 0$.

\medskip\noi
{\bf Proof.} From (4.32)-(4.34) we obtain
\beq            %4.37
\nhq {R_{MNPQ}}[g] {R^{MNPQ}}[g] \sim A_1
  [C + \eta_{ab} x^a x^b]^{-2 - 2 \alpha}
\ee
as $C + \eta_{ab} x^a x^b \to + 0$, where
\bearr             %4.38
A_1 =  (\alpha {-}1 )^2 (N_0 {-}1) C^2 [16 + 2 (N_0{-}2)(1{-} \alpha)^2]\nnn
+ 2 N \alpha^2 C^2
[2 + (N {-} 1) \alpha^2 + 2 (N_0 {-} 1)(1{-} \alpha)^2] > 0.\nnn
\ear
Then (4.36) follows from (4.37), (4.38) and $\alpha > 0$.

Thus the solution  (4.28) with
a non-Euclidean signature $\eta = (\eta_{ab}) \neq (\delta_{ab})$
and $C \neq 0$  cannot be extended  to the manifold (4.23).

For $N_0 {=}4$, $\sum_{i=1}^{n} N_i = 6$,
$\eta = \pm {\rm diag} ( - 1, 1, 1, 1)$, we get an ${O}(1,3)$-symmetric
solution in 10-dimensional gravity with a pseudo-Euclidean
conformally flat 4-dimensional section
\bearr            %4.39
g = [C \pm x^2]^{1/2}
\biggl\{ \frac{ - dx^{0} \otimes dx^{0} +
d \vec{x} \otimes d \vec{x}}{ \pm C  + x^2}      \nnn
\cm \cm  + \sum_{i=1}^{n} \frac{\lambda_i}{N_0 - 2} g^i  \biggr\},
\ear
where $ x^2 = - (x^{0})^{2} + (\vec{x})^{2}$.

\medskip\noi
{\bf Remark 4.} The  ``Euclidean" solution (4.35) with $C =1$
may be also written in the form
\bearr           %4.40
g = (\cosh y)^{2 \alpha} \biggl\{ dy \otimes dy \nnn
\cm +  \tanh^2 y d \Omega^2_{N_0 -1}
+ \sum_{i=1}^{n} \frac{\lambda_i}{N_0 - 2} g^i  \biggr\},
\ear
where $\sinh y = r$  and $\alpha = 1 - 2/N_0$). The
$N_0$-dimensional section of (4.40) contains a
"sigar-type" metric multiplied by a conformal factor:
\beq            %4.41
g_s = (\cosh y)^{2 \alpha}
\{ dy \otimes dy +  \tanh^2 y \ d \Omega^2_{N_0 -1} \}.
\ee

\medskip\noi
{\bf De Sitter membrane.}  Let $n =1$ and
\beq            %4.42
\nq g^{1} = g(dS^{N_1}) =
- dt \otimes dt +  \frac{\cosh^2 (Ht)}{H^2}
d \Omega^2_{N_1 -1}
\ee
be the $N_1$-dimensional de Sitter metric, where  $N_1$ is defined
in (4.16) and
\beq            %4.43
H^2 = \frac{N_0 - 2}{N_1 - 1} =  \frac{(N_0 - 2)^2}{3 N_0 - 2}.
\ee
The metric (4.42) satisfies the relation
\beq            %4.44
{\rm Ric} \ [ g(dS^{N_1})] = (N_0 - 2) \ g(dS^{N_1}),
\ee
and hence the metric
\bearr           %4.45
g = [C + r^2]^{\alpha}
\biggl\{ \frac{dr \otimes dr + r^2 \Omega^2_{N_0 -1}}{C + r^2}\nnn
\cm  - dt \otimes dt +  \frac{\cosh^2 (Ht)}{H^2}
          d \Omega^2_{N_1 -1} \biggr\},
\ear
($\alpha = 1 - 2/N_0$)  satisfies the Einstein equations. The metric (4.45)
describes a spherically symmetric nonsingular de Sitter membrane solution.

\medskip\noi
{\bf The curvature-splitting trick.} The solution
(4.28) with $n$ internal spaces may be obtained from the one with $n=1$
by so-called "curvature-splitting"  trick \cite{GIM2}.
Let us consider a set of $k$  Einstein manifolds $({\cal M}_i, h^i )$
of nonzero curvature, i.e.
\beq           %4.46
{\rm Ric}\ (h^i )= \mu_i h^i ,
\eeq
where $\mu_i \neq 0$  is a real constant, $i = 1, ..., k$.
Let $\mu \neq 0$ be a real number. Then
\beq           %4.47
h= \sum_{i=1}^{k} \frac{\mu_i}{\mu} h^i
\eeq
is an Einstein metric, (correctly) defined on
\beq           %4.48
{\cal M} =  {\cal M}_1 \times \ldots \times {\cal M}_k
\eeq
and satisfying
\beq           %4.49
{\rm Ric}\ (h)= \mu h.
\eeq
Indeed,
\bearr           %4.50
{\rm Ric}\ (h)= \sum_{i=1}^{k} {\rm Ric}\
     \Bigl(\frac{\mu_i}{\mu} h^i \Bigr) \nnn
= \sum_{i=1}^{k} {\rm  Ric}(h^i ) = \sum_{i=1}^{k}\mu_i h^i  = \mu h.
\ear
(Here we have simplified the notations according to Remark 1.)

\psection{The case $N_0 =2$}

Consider now the exceptional case $N_0 = 2$. In this
case the action (2.12) reads  (we put here $\kappa^{2}_0 = 1$)
\bearr           %5.1
S = S_{\sigma}[g^0 , \gamma,\phi] \nnn
= \frac{1}{2}
     \int_{M_0} d^{2}x \sq {g^0}
     \exp \biggl(\sum_{i =1}^{n} N_i \phi^i\biggr)\biggl\{ {R}[g^0 ]\nnn
\cm -  \bar{G}_{ij} (\p \phi^i) (\p \phi^j)
          + 2 (\p \gamma) \sum_{j =1}^{n} N_j \p \phi^j  \nnn
\cm + \sum_{i=1}^{n} \lambda_{i} N_i \e^{-2 \phi^i + 2 \gamma} -
          2 \Lambda \e^{2 \gamma}\biggr \},
\ea
where $\bar{G}_{ij}$
is the cosmological minisuperspace metric (3.49).
>From (5.1) we see that the midisuperspace metric
crucially depends upon the choice of $\gamma$.
For  $\gamma = 0$ we get from (5.1) the action
with a conformally  flat  midisuperspace metric of pseudo-Euclidean
signature
\bearr           %5.2
S  = \frac{1}{2}
\int_{M_0} d^{2}x \sq {g^0}
\exp\biggl(\sum_{i =1}^{n} N_i \phi^i\biggr) \biggl\{ {R}[g^0 ]  \nnn
\nq -  \bar{G}_{ij}
(\p_{\mu} \phi^i) (\p_{\nu} \phi^j)  {g^0}^{ \mu \nu}
+ \sum_{i=1}^{n} \lambda_{i} N_i \e^{-2 \phi^i } \! -2 \Lambda \biggr \}.
\ear

Another choice of the conformal frame parameter
\beq            %5.3
\gamma = - \frac{1}{2}  \sum_{i=1}^{n} N_i \phi^i
\ee
leads us to the action
\bearr           %5.4
S =  \frac{1}{2}
\int_{M_0} d^{2}x \sq {g^0}
\exp\biggl(\sum_{i =1}^{n} N_i \phi^i\biggr) \biggl\{ {R}[g^0 ]  \nnn
     -  \sum_{i =1}^{n} N_i
     (\p_{\mu} \phi^i) (\p_{\nu} \phi^i)  {g^0}^{\mu \nu}  \nnn
+ \biggl(\sum_{i=1}^{n} \lambda_{i} N_i
     \e^{-2 \phi^i} - 2 \Lambda \biggr)
\exp\biggl(- \sum_{i =1}^{n} N_i \phi^i\biggr) \biggr\},
\ear
with a Euclidean conformally flat midisuperspace metric.
Note that in \Ref{RZ} the action (5.2) was reduced
to a ``string-like" form (for $n = 1$ see, for example, \cite{Ca}).

\psection{Appendix}

\psubsection{Riemann tensor.}

Here we consider the metric
\beq           %6.1
g=  \bar{g}^0  +
\sum_{i=1}^{n} \e^{2\phi^i (x)} g^i ,
\eeq
defined on the manifold (2.1), where the metrics $\bar{g}^0 $ and
$g^i $ are defined on $M_0$ and $M_i$ respectively, $i=1,\ldots,n$.
The nonzero components of the Riemann tensor corresponding
to (6.1) are
\bear           %6.2 -6.4
  R_{\mu \nu \rho \sigma }[g] \eql
R_{\mu \nu \rho \sigma }[\bar{g}^0 ], \\
  R_{\mu m_i \nu n_i }[g]    \eql - R_{m_i \mu \nu n_i }[g] =
                                     - R_{\mu m_i n_i \nu }[g]      \nn
  = R_{m_i \mu n_i \nu }[g] \eql - \e^{2 \phi^i} g^i _{ m_i n_i}
[\btd_{\mu}[\bar{g}^0 ](\p_{\nu} \phi^i)  \nnn
                     +(\p_{\mu} \phi^i) (\p_{\nu} \phi^i)],  \\
   R_{m_i n_j p_k q_l }[g]  \eql \e^{2 \phi^i}
\delta_{ij} \delta_{kl} \delta_{ik} R_{m_i n_i p_i q_i }[g^i ] \nnn
\nqq\nqq\nq  + \e^{2 \phi^i + 2 \phi^j}
  \bar{g}^{0\ \mu \nu} (\p_{\mu} \phi^i) (\p_{\nu} \phi^j)
[ \delta_{il} \delta_{jk}   g^i _{ m_i q_i} g^{j}_{ n_j p_j} \nnn
\nq\nq\nq -\delta_{ik} \delta_{jl} g^i _{ m_i p_i} g^j _{ n_j q_j} ],
\ear
where the indices $\mu, \nu, \rho, \sigma$ correspond to $M_0$,
$m_i$, $n_i$, $p_i$, $q_i$  to $M_i$; $i,j,k,l= 1, \ldots, n$,
${\btd}[g^0 ]$ is a covariant derivative with respect to $g^0 $.

The relations (6.2)-(6.4) may be obtained from the following
relations for the nonzero components of the Christophel-Schwarz symbols:
\bear           %6.5 -6.8
  \Gamma^{\mu}_{ \nu \rho }[g] \eql
\Gamma^{\mu}_{ \nu \rho }[\bar{g}^0 ], \\
  \Gamma^{m_i}_{n_i \nu }[g] \eql \Gamma^{m_i}_{\nu n_i }[g]
= \delta ^{m_i}_{n_i} \p_{\nu} \phi^i,          \\
  \Gamma^{\mu}_{ m_i n_i }[g] \eql
- \bar{g}^{0\ \mu \nu} (\p_{\nu} \phi^i)
          \e^{2 \phi^i}  g^i _{ m_i n_i},  \\
  \Gamma^{m_i}_{ n_i p_i }[g] \eql
\Gamma^{m_i}_{ n_i p_i }[g^i ].
\ear

\psubsection{Riemann tensor squared.}

We denote the squared Riemann tensor by
\beq             %6.9
{I}[g] \equiv {R_{MNPQ}}[g] {R^{MNPQ}}[g].
\eeq
As follows from \eqs (6.2)-(6.4), for the metric (6.1)   \cite{IM4}
\bearr           %6.10
\nq {I}[g] = {I}[\bar{g}^0 ] +
       \sum_{i=1}^{n} \{ \e^{-4 \phi^{i}} {I}[g^{i}]
- 4 \e^{-2 \phi^i} {U}[\bar{g}^0 , \phi^i] {R}[g^i ] \nnn
\cm - 2 N_i {U^2}[\bar{g}^0 , \phi^i]  +
4 N_i {V}[\bar{g^0 }, \phi^i] \} \nnn
\cm +
\sum_{i,j =1}^{n} 2 N_i N_j [\bar{g}^{(0), \mu \nu} (\p_{\mu} \phi^i)
\p_{\nu} \phi^j]^2 ,
\ear
where
${R}[g^i ]$ is the scalar curvature
of $g^i $  and  $N_i = \dim  M_i$, $i = 1, \ldots, n$.  In (6.10)
\bear            %6.11 -12
{U}[g,\phi] \al \equiv \al g^{MN} (\p_{M} \phi) \p_{N} \phi, \\
{V}[g,\phi] \al \equiv \al g^{M_{1}N_{1}} g^{M_{2}N_{2}} \times \nnn
\times [\btd_{M_1}(\p_{M_2} \phi)
                    +  (\p_{M_1} \phi) \p_{M_2} \phi] \times    \nnn
\times [\btd_{N_1}(\p_{N_2} \phi)  +  (\p_{N_1} \phi) \p_{N_2} \phi],
\ear
where $\btd = {\btd}[g]$ is a covariant derivative with respect to $g$.

\psubsection{The cosmological case}

Consider now the special case of (6.10)  with
$M_0 = (t_1, t_2)$, $t_1 < t_2$.
Thus we consider the metric
\beq            %6.13
g_c = - {B}(t) dt \otimes dt +  \sum_{i=1}^{n} {A_{i}}(t) g^i ,
\ee
defined on the manifold
\beq            % 6.14
M = (t_1, t_2) \times M_{1} \times \ldots \times M_{n}.
\ee
and  ${B}(t), {A_{i}}(t) \neq 0$, $i = 1, \ldots, n$.

>From  (6.11) we obtain the Riemann tensor squared for the metric
(6.13) \cite{IM3,IM4}
\bearr           %6.15
{I}[g_c] =
 \sum_{i=1}^{n} \Bigl\{ A_{i}^{-2} {I}[g^i ] + A_{i}^{-3} B^{-1}
\dot{A}_{i}^{2} {R}[g^i ]  \nnn
\quad  - \frac{1}{8}N_{i} B^{-2} A_{i}^{-4} \dot{A}_{i}^{4}
     +\frac{1}{4} N_{i} B^{-2}  \bigl(2 A_{i}^{-1} \ddot{A}_{i} \nnn
\quad     - B^{-1} \dot{B} A_{i}^{-1} \dot{A}_{i}
           - A_{i}^{-2} \dot{A}_{i}^{2}\bigr)^{2} \Bigr\}    \nnn
\quad + \frac{1}{8}B^{-2} \biggl[\sum_{i=1}^{n} N_{i} (A_{i}^{-1}
               \dot{A}_{i})^{2}\biggr]^{2}.
\ear

\psubsection{Conformal transformation}

We present for convenience the well-known relations \cite{Kr}
\bearr           %6.16-18
\e^{-2 \gamma} R_{\mu \nu \rho \sigma }[\e^{2 \gamma} g^0 ]
  =    R_{\mu \nu \rho \sigma }[g^0 ]   \nnn
     \qquad + Y_{\nu \rho} g^0 _{\mu \sigma } -
Y_{\mu \rho} g^0 _{\nu \sigma } -
Y_{\nu \sigma}  g^0 _{\mu \rho } +
Y_{\mu \sigma}  g^0 _{\nu \rho },  \\
\lal R_{\mu \nu}[\e^{2 \gamma} g^0 ]
= R_{\mu \nu}[ g^0 ] + (2 {-} N_0) Y_{\mu \nu}
  - g^0 _{\mu \nu} {g^0}^{ \rho \tau}  Y_{\rho \tau}, \nnn \\
\lal  \btu[\e^{2 \gamma} g^0 ] =
\e^{- 2 \gamma} \{ \Delta_0+
(N_0 {-} 2) {g^0}^{ \mu \nu} (\p_{\mu} \gamma) \p_{\nu} \} \nnn
\ea
where,  as in Subsec.\,2.1,  the metric $g^0 $  is defined
on $M_0$, $\dim M_0 = N_0$, $\Delta_0$ is the
Laplace-Beltrami operator on $M_0$  and
\beq            %6.19
Y_{\mu \nu} = \gamma_{; \mu \nu} -  \gamma_{\mu } \gamma_{\nu }
+ \half g^0_{\mu \nu }  \gamma_{\rho } \gamma^{\rho}.
\ee

\Acknow
{The authors are grateful to K.A. Bronnikov, A.I. Zhuk and V.A. Berezin
for useful discussions. We are also grateful to the organizers and
participants of the 9th Russian gravitational conference in Novgorod and
the School at Nordfjordeid (Norway) where the results of this paper were
reported.

This work was supported in part by the Russian State Committee for
Science and Technology, Russian Fund for Basic Research and
CNRS, France (for V.N.M.).}

\end{document}